\documentclass[12pt,preprint]{aastex}

\shorttitle{First Doppler peak}
\shortauthors{Gonz\'{a}lez-D\'{\i}az}

\begin{document}

\title{Quintessence and the first Doppler peak}

\author{Pedro F. Gonz\'{a}lez-D\'{\i}az\altaffilmark{1}}
\affil{Instituto de Matem\'{a}ticas y F\'{\i}sica Fundamental, Consejo Superior de\\
Investigaciones Cient\'{\i}ficas, Serrano 121, 28006 Madrid, Spain }

\altaffiltext{1}{Centro de F\'{\i}sica "M.A. Catal\'{a}n", E-mail: p.gonzalezdiaz@imaff.cfmac.csic.es}

\begin{abstract}
By using a tracking quintessence model we obtain that the
position of the first Doppler peak in the spectrum of CMB
anisotropies only depends on the topology of the universe,
$\Omega_k$, for any value of the ratio
$\Omega_{\Lambda}/\Omega_{M}$, so that such a dependence is
perfectly valid in the range suggested by supernova
observations.
\end{abstract}


\section{Introduction}

Recent experiments and observations have already opened up what
one can actually call the era of precision cosmology. It is now
hoped that next years will see rather impressive advances
leading not just to the determination of key cosmological
parameters with an accuracy not even dreamed five years ago, but
also to unprecedented scrutinies on fundamental aspects of
particle and field theories. In particular, the connection
between recent Boomerang (de Bernardis et al. 2000) and Maxima
(Hanany et al. 2000) experiments with theory is twofold. They
are linked through predictions from either inflationary (and
perhaps cosmic string) models by using fundamental particle
physics arguments (Hu and White 1996) or from general more or
less standard cosmological scenarios which may or may not
include some new concepts such as quintessence (Cadwell, Dave
and Steinhardt 1998). The aim of this paper will concentrate at
a particular aspect of CMB anisotropy measurements: the
dependence of the position of the first Doppler peak on the
values of the relevant cosmological parameters within the realm
of a quintessence model. Kamionkowski, Spergel and Sugiyama
(1994) originally derived the simple relation that the position
of the first Doppler peak $\ell\sim 200\Omega_0^{-1/2}$, where
$\Omega_0=\Omega_M+\Omega_{\Lambda}$, which has been the subject
of some controversy. While Frampton, Ng and Rohm (1998) have
recently derived a similar dependence using a simple
quintessential model, Weinberg (2000) has argued that this
formula is not even a crude approximation when $\Omega_M$ is
smaller than $\Omega_{\Lambda}$. For a very recent discussion on
quntessential model of CMB anisotropies see Bond et al. (2000).
Assuming the holding of the cosmic triangle condition, we obtain
in this paper that $\ell$ only depends on the topology of the
universe, so confirming the original proposal by Kamionkowski,
Spergel and Sugiyama (1994).

\section{Calculation of $\ell$}

In order to derive an expression relating the position of the
first Doppler peak with the scalar field potential of
quintessence models, we choose a general tracking model with
time-dependent parameter $\omega(t)$ for the state equation
(Zlatev, Wang and Steinhardt 1999). This kind of models can be
related to particle physics and may solve the so-called cosmic
coincidence problem (Steinhardt 1997). One typically considers
tracking quintessence fields $\phi$ for a Ratra-Peebles
potential $V(\phi)\propto\phi^{-\alpha}$ with a given constant
parameter $\alpha$ (Ratra and Peebles 1988). During its cosmic
evolution, the equation of state parameter $\omega$ passes
through several distinct regimes (Brax, Martin and Riazuelo
2000), including first a kinetic regime lasting until $z\sim
10^{20}$ with $\omega=+1$, then the transition and potential
regimes chracterized by $\omega=-1$, to finally reach the proper
tracking regime at $z\sim 10^{15}$, where
$\omega=(\alpha\omega_B -2)/(\alpha+2)$ (with $\omega_B=1/3$ for
radiation dominated universe and $\omega_B=0$ for matter
dominated universe). On this regime, there exists a particular
solution for the scalar field, $\phi=\phi_0
\eta^{4/(\alpha+2)}$ (in which $\eta=\int dt/R(t)$ is the
conformal time, with $R(t)$ the scale factor) which is an
attractor able to solve the cosmic coincidence problem.

For $\alpha\neq 0$ a change of $\omega$ should then be expected
when the universe becomes matter dominated once the the surface
of last scattering is overcome. Since the tracking regime is
characterized by $\omega=c_{s}^2$ ($c_{s}^2$ being the sound
velocity defined as $(dp_{\phi}/d\eta)/(d\rho_{\phi}/d\eta)$
(Brax, Martin and Riazuelo 2000)), parameter $\omega$ becomes a
constant given by $\omega=-2/(\alpha+2)$ when matter dominates.
It has been recently argued that $\alpha\leq 2$ (Balbi et al.
2001), so that if we choose for definiteness $\alpha=1$, then
$\omega=-2/3$ for the vacuum quintessence field. In this case,
the Ratra-Peebles potential becomes
\begin{equation}
V(\phi)\propto \phi^{-1} .
\end{equation}
Since the corresponding tracking solution,
$\phi\propto\eta^{4/3}$, would then correspond to a constant
parameter $\omega < 0$ for the quintessence field, it should
satisfy the constraint equation derived from the corresponding
conservation laws and cosmological field equations (Di Pietro
and Demaret 1999, Gonz\'{a}lez-D\'{\i}az 2000). In terms of the
cosmological parameters $\Omega_i$, $i=M,\phi,\Lambda, k$, and
the quintessence field potential $V(\phi)$, this constraint
equation can be written as (Gonz\'{a}lez-D\'{\i}az 2000)
\[\left(\frac{V'}{V_0 '}\right)^2=\]
\begin{equation}
\Omega_M\left(\frac{V}{V_0}\right)^{(\omega+2)/(\omega+1)}+
\Omega_{\phi}\left(\frac{V}{V_0}\right)^{2}
+\Omega_k\left(\frac{V}{V_0}\right)^{(3\omega+1)/[3(\omega+1)]}
+\Omega_{\Lambda}\left(\frac{V}{V_0}\right)^{(3\omega+4)/[3(\omega+1)]},
\end{equation}
where $'\equiv d/d\phi$, the subscript 0 means current value and
$V_0/v_0 '=\pm\sqrt{3\Omega_{\phi}/(1+\omega)}$. Eqn. (2)
corresponds to the generalized quintessence model with negative
constant parameter for the state equation recently suggested
(Gonz\'{a}lez-D\'{\i}az 2000). Besides the contributions from the
topological curvature, $k$, and gravitationally observable mass,
$M$, this model distinguishes two essentially distinct
contributions from vacuum energy: a varying cosmological term
with positive energy density $\Omega_{\Lambda}$ assumed to
satisfy the conservation law $8\pi
G\rho_{\Lambda}=\Lambda_0(R_0/R)$ at sufficiently small
redshifts, and a quintessence negative energy density such that
while $\Omega_{\phi}=8\pi G\rho_{\phi}/3H_0^2$ ($H_0$ being the
current value of the Hubble constant) is always negative,
$\Omega_v\equiv\Omega_{\Lambda}+\Omega_{\phi}$ is always
positive, so satisfying the Ford-Roman's quantum interest
conjecture (Ford and Roman 1999).

Adapting then the relation between the position of the first
Doppler peak $\ell$ and the cosmological parameters $\Omega_i$
first derived by Frampton, Ng and Rohm (1998)to our generalized
model, we can finally obtain a general relation between $\ell$
and the quintessence potential $V(\phi)$ of the form:
\[\ell=
\frac{6\pi(1+\omega)V_0}{(3\omega+1)\sqrt{|\Omega_k|}V_0
'}\times\]
\begin{equation}
\frac{d}{d\phi}
\left.\left[
\left(\frac{V}{V_0}\right)^{(3\omega+1)/[6(\omega+1)]}\right]\right|_{z=Z_R}
S\left\{\frac{\sqrt{|\Omega_k|}V_0 '}{3(\omega+1)v_0}
\int_{\phi(z=0)}^{\phi(z=z_R)}
\frac{d\phi}{\left(\frac{V}{V_0}\right)^{(3\omega+1)/[6(\omega+1)]}}\right\}
,
\end{equation}
where $z_R=1100$ is the redshift at recombination, we have used
the relation (Di Pietro and Demaret 1999)
\begin{equation}
\frac{V}{V_0}=\left(\frac{R_0}{R}\right)^{3(\omega+1)}=(1+z)^{3(\omega+1)},
\end{equation}
and $S(x)=1$ for $\Omega_k=0$, $S(x)=\sin(x)$ for $\Omega_k <0$
and $S(x)=\sinh(x)$ for $\Omega_k>0$. In the case of interest
$\omega=-2/3$, where we obtain both topological and dynamical
acceleration, Eqn. (3) reduces to
\begin{equation}
\ell=\left.\frac{2\pi
V_0}{\sqrt{|\Omega_k|}V_0 '}
\frac{d}{d\phi}
\left[\left(\frac{V}{V_0}\right)^{-1/2}\right]\right|_{z=z_R}
S\left\{\frac{\sqrt{|\Omega_k|}V_0 '}{V_0}
\int_{\phi(z=0)}^{\phi(z=z_R)}d\phi\left(\frac{V}{V_0}\right)^{1/2}\right\}
,
\end{equation}
and there is a solution to the constraint equation (2) given by
\begin{equation}
V=V_0\left\{\sinh\left[\beta_0(\phi-\phi_0)\right]+\xi_0\right\}^{-1},
\end{equation}
where
\begin{equation}
\beta_0=\pm\frac{1}{3}\sqrt{\frac{\Omega_{\Lambda}+\Omega_{\phi}}{\Omega_{\phi}}}
=\pm\frac{1}{3}\sqrt{\frac{\Omega_v}{\Omega_{\phi}}}
\end{equation}
\begin{equation}
\xi_0=-\frac{\Omega_k}{2\Omega_v} .
\end{equation}
In this case, the cosmological parameters must satisfy the
relations
\begin{equation}
\Omega_v=\frac{1}{4} \left[1-\Omega_k +\sqrt{1-2\Omega_k
-\Omega_k^2}\right]\geq 0
\end{equation}
\begin{equation}
\Omega_M=\frac{1}{4}  \left[3(1-\Omega_k)-\sqrt{1-2\Omega_k
-\Omega_k^2}\right]> 0 .
\end{equation}
Together with the triangle equation
$\Omega_v+\Omega_M+\Omega_k=1$, Eqns. (9) and (10) will restrict
the values that the cosmological parameters may take on. On the
other hand, we note from Eqns. (4) and (6) that for large $z$
and $\Omega_k\sim 0$ one obtains $\phi\propto\eta^2\propto R$
and $V\propto\phi^{-1}\propto R^{-1}$; that is, while the
potential (6) can be consistently interpreted as a
generalization for smaller values of $z$ from the Ratra-Peebles
potential, the attractor solution $\phi\propto R^{4/3}$ must
necessarily change into $\phi\propto R$ as the universe becomes
matter dominated.

Performing the derivative and integration in Eqn. (5) after
inserting solution (6), we finally obtain for the position of
the first Doppler peak when $\omega=-2/3$
\begin{equation}
\ell=\pi\sqrt{\frac{\Omega_v(1+z_R)}{|\Omega_k|} \left[1
+\left(\frac{1}{1+z_R}-\frac{|\Omega_k|}{2\Omega_v}\right)^2\right]}
\left.T\left\{\sqrt{\frac{|\Omega_k|}{\Omega_v\varphi_0}}
F\left[\varphi(z), r_0\right]\right|_{0}^{z_R}\right\} ,
\end{equation}
where $F[\varphi,r_0]$ is the elliptic integral of the first
kind (Abramowitz and Stegun 1965),
\begin{equation}
\varphi=
\arccos\left(\frac{\varphi_0-\frac{1}{1+z}}{\varphi_0+\frac{1}{1+z}}\right)
,
\end{equation}
\begin{equation}
r_0=
\left(\frac{\frac{|\Omega_k|}{2\Omega_v}+\varphi_0}{2\varphi_0}\right)^{1/2},
\end{equation}
\begin{equation}
\varphi_0=\sqrt{1+\left(\frac{|\Omega_k|}{2\Omega_v}\right)^2} ,
\end{equation}
and $T(x)=1$ for $\Omega_k=0$ and, unlike the function $S(x)$,
$T(x)=\sin(x)$ both for the closed and open cases. $\Omega_k$
can run between two extreme values, such that
\begin{equation}
-(1+\sqrt{2})\leq\Omega_k\leq\sqrt{2}-1 .
\end{equation}

A plot for the position of the first Doppler peak, $\ell$,
against the parameter $\Omega_k$ is given in Fig. 1. It can be
seen that the maximum value for $\ell$ is reached when $\ell$
approaches the value 190 at the flat case $\Omega_k=0$,
decreasing slowly therefrom as $|\Omega_k|$ increases, a little
more steeply for $\Omega_k
>0$ than for $\Omega_k <0$. This seems to quite reasonably
reproduce the results provided by Boomerang, Maxima and previous
(Netterfield et al. 1997, Vollaek et al. 1997) experiments.
Although uncertainties about our results for $\omega=-2/3$ must
come from the possibility of having values of $\alpha$ other
than just unity (provided they are on the interval (0,2), or
small deviation from time-independence of $\omega$ during cosmic
evolution after recombination, they seems to strongly suggest a
nearly flat topology for the universe.

We note that CBM anisotropies and cosmic acceleration can be
easily related in our model. To see this, let us consider
solution (6) for the the flat case, i.e.
\[V=V_0
\left\{\sinh\left[\pm\sqrt{\frac{\Omega_M}{\Omega_{\phi}}}
(\phi-\phi_0)\right]\right\}^{-1} ,\] with
\[\Omega_v=\Omega_M=\frac{1}{2} .\]
This allows us to also construct a plot for the luminosity
distance versus redshift by using the expression (Gonz\'{a}lez-D\'{\i}az
2000)
\begin{equation}
D_L H_0=\sqrt{2}(1+z)
\left.\left\{F\left[\arccos\left(\frac{1-\frac{1}{1+z'}}{1+\frac{1}{1+z'}}\right),
45^{\circ}\right]\right|_{0}^{z}\right\} .
\end{equation}
One can see readily that our flat solution in fact gives rise to
a $5\log D_L H_0 - z$ plot with a nearly straight line between
$z\simeq 0.01$ and $z\simeq 0.5$ which appears to slightly
accelerate thereafter, the full $5\log D_L H_0$ that corresponds
to the $z$-interval of presently available type Ia supernova
observations, $[\simeq(0.01-1)]$, being around 12. This
accelerating behaviour for the universe conforms quite well the
data obtained from distant supernova Ia (Perlmutter et al. 1999,
Reiss et al. 1998). It is worth noticing that though our
universe model is dynamically accelerating it still is
topologically uniform as $q_0=\frac{1}{2}(\Omega_M-\Omega_v)$
exactly vanishes in the flat case. As one separates from
flatness one model produces topologically decelerating
scenarios.

\section{Conclusions}

Consistent solutions to the constraint equation for the
quintessence potential have been obtained which all correspond
to particular fixed sets of values for $\Omega_i$,
$i=k,M,\phi,\Lambda$, in such a way that the resulting value of
the position of the first Doppler peak $\ell$ becomes
automatically fixed once just one of the parameters $\Omega_k$,
$\Omega_M$ or $\Omega_v$ is fixed, but does not depend on the
ratio $\Omega_{\phi}/\Omega_{\Lambda}<0$. One can also conclude
that $\ell$ acquires a maximum value which fits fairly well
experimental results at $\Omega_k=0$ (i.e.
$\Omega_M=\Omega_v=1/2$). As the topology of the universe
separates from flatness ($\Omega_M,\Omega_v >1/2$ if $\Omega_k
<0$, or $\Omega_M,\Omega_v <1/2$ if $\Omega_k >0$), $\ell$
always decreases from its maximum value. Since
$\Omega_{\phi}<0$, from the condition that $V/V_0$ be real it
follows that the quintessence field $\phi$ is pure imaginary.
This amounts to the interpretation that the classical solution
is axionic, and hence the quintessence field can be taken as a
genuine component of dark matter in such a way that one can
define for the gravitationally observable mass of the universe
the quantity $\Omega_M '=\Omega_M+\Omega_{\phi}>0$, with
$\Omega_v >0$. Thus, one can always adjust our results to the
currently favoured combination $\Omega_M '=0.3$, $\Omega=0.7$
for any $\Omega_k$.

We finally remark that the results obtained in this paper refer
only to the case of a quintessence parameter for state equation
$\omega=-2/3$ (i.e. $\alpha=1$). It is expected that results
even more adjusted to observation (i.e. closer to 200) can also
be obtained within the model of this paper by slightly shifting
$\omega$ towards more realistic smaller values approaching -0.8.

\acknowledgments

For helpful comments, the author thanks C. Sig\"{u}enza. This
research was supported by DGICYT under research project No.
PB97-1218.

\clearpage

\begin{figure}
\scalebox{.7}{\plotone{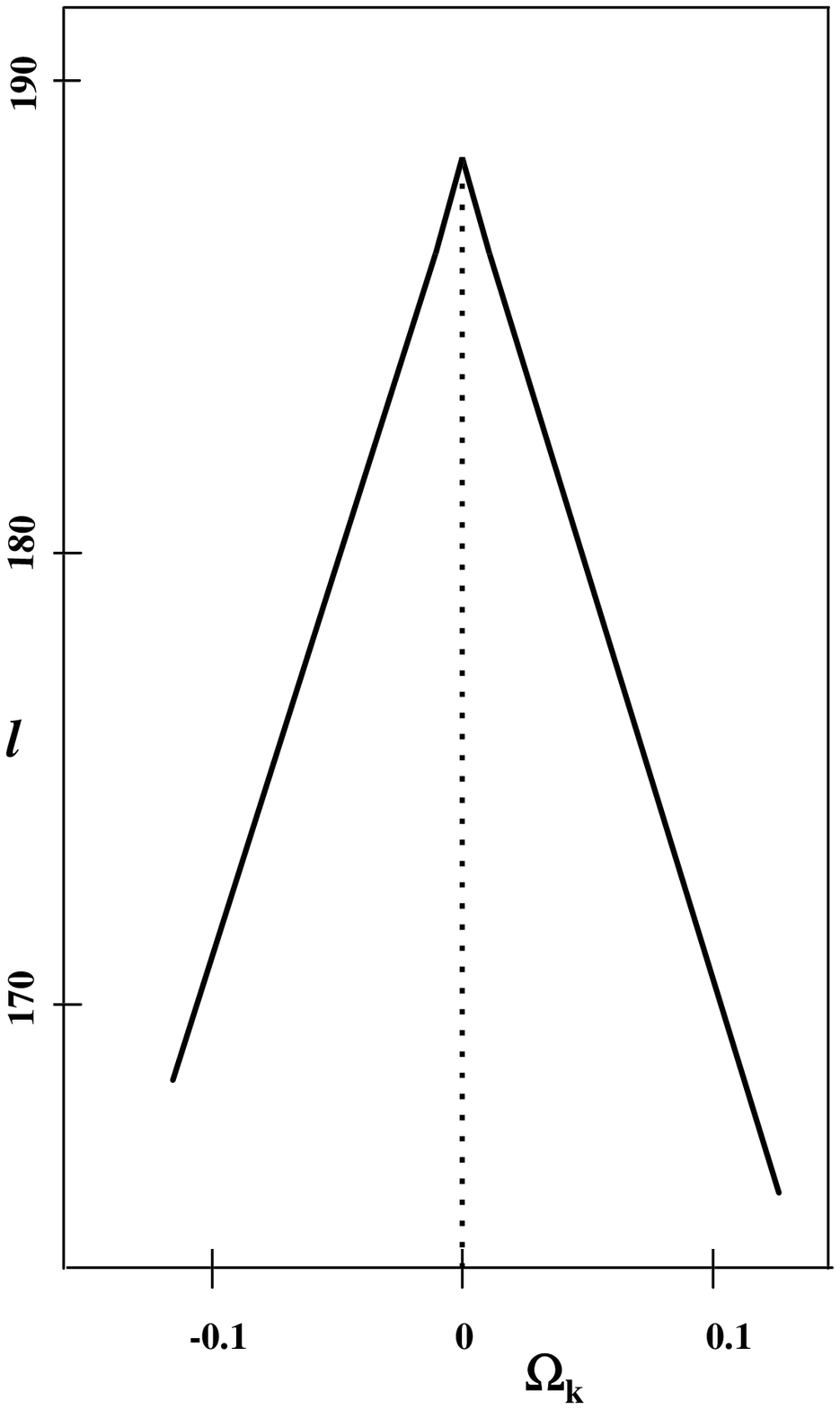}}
\caption{Dependence of the first Doppler peak on the curvature of the universe
for a quintessence model with constant parameter $\omega=-2/3$.
}
\label{rad}
\end{figure}

\end{document}